\documentclass[aoas,MSNbibl,nameyear,seceqn,dvips]{arximspdf}
\usepackage{graphicx}

\doi{10.1214/12-AOAS591} 
\volume{6}
\issue{4}
\pubyear{2012}
\firstpage{1430}
\lastpage{1451}

\makeatletter

\newcommand{\bbeta}{\bolds{\beta}}
\newcommand{\bA}{\mathbf{A}}
\newcommand{\bX}{\mathbf{X}}
\newcommand{\bs}{\mathbf{s}}
\newcommand{\bu}{\mathbf{u}}
\newcommand{\bv}{\mathbf{v}}
\newcommand{\iid}{\stackrel{\mathrm{i.i.d.}}{\sim}}
\newcommand{\indep}{\stackrel{\mathrm{indep}}{\sim}}
\newcommand{\calR}{\mathcal{R}}
\newcommand{\calD}{\mathcal{D}}
\newcommand{\calS}{\mathcal{S}}
\newcommand{\Frechet}{\mbox{Fr\'{e}chet}}
\newcommand{\Matern}{ \mbox{Mat\'{e}rn}}

\makeatother

\begin{document}
\begin{frontmatter}

\title{A hierarchical max-stable spatial model for extreme precipitation\thanksref{T1}}
\runtitle{A max-stable model for extreme precipitation}
\thankstext{T1}{Supported in part by the NSF
(DMS-07-06731, Reich and DMS-09-14906, Shaby), the US Environmental
Protection Agency (R835228, Reich), the National Institutes of Health
(5R01ES014843-02, Reich), as well as the Statistics and Applied
Mathematical Sciences Institute (SAMSI).}

\begin{aug}
\author[a]{\fnms{Brian J.} \snm{Reich}\corref{}\ead[label=e1]{brian\_reich@ncsu.edu}}
\and
\author[b]{\fnms{Benjamin A.} \snm{Shaby}\ead[label=e2]{bshaby@stat.berkeley.edu}}
\runauthor{B. J. Reich and B. A. Shaby}
\affiliation{North Carolina State University
and University of California---Berkeley}
\address[a]{Department of Statistics\\
North Carolina State University\\
Raleigh, North Carolina 27560\\
USA\\
\printead{e1}}

\address[b]{Department of Statistics\\
University of California---Berkeley\\
Berkeley, Berkeley 94720\\
USA\\
\printead{e2}}
\end{aug}

\received{\smonth{11} \syear{2011}}
\revised{\smonth{8} \syear{2012}}

%
\begin{abstract}
Extreme environmental phenomena such as major precipitation events
manifestly exhibit spatial dependence. Max-stable processes are a class
of asymptotically-justified models that are capable of representing
spatial dependence among extreme values. While these models satisfy
modeling requirements, they are limited in their utility because their
corresponding joint likelihoods are unknown for more than a trivial
number of spatial locations, preventing, in particular, Bayesian
analyses. In this paper, we propose a new random effects model to
account for spatial dependence. We show that our specification of the
random effect distribution leads to a max-stable process that has the
popular Gaussian extreme value process (GEVP) as a limiting case. The
proposed model is used to analyze the yearly maximum precipitation from
a regional climate model.
\end{abstract}

%
\end{frontmatter}

\section{Introduction}\label{sintro}
Spatial statistical techniques are crucial for accurately quantifying
the likelihood of extreme events and monitoring changes in their
frequency and intensity. Extreme events are by definition rare,
therefore, estimation of local climate characteristics can be improved
by borrowing strength across nearby locations. While methods for
univariate extreme data are well developed, modeling
spatially-referenced extreme data is an active area of research.
Max-stable processes [\citet{dehaan-2006a}] are the natural
infinite-dimensional generalization of the univariate generalized
extreme value (GEV) distribution. Just as the only limiting
distribution of the scaled maximum of independent univariate random
variables is the GEV, the scaled maximum of independent copies of any
stochastic process can only converge to a max-stable process.
Max-stable process models for spatial data may be constructed using the
spectral representation of \citet{dehaan-1984a}. Max-stable processes
built from this representation were first used for spatial analysis by
\citet{smith-1990a}. Since then, a handful of subsequent spatial
max-stable process models have been proposed, notably that of \citet
{schlather-2002a} and \citet{kabluchko-2009a}, who proposed a more
general construction that includes several other known models as
special cases. Applications of spatial max-stable processes include
\citet{coles-1993a}, \citet{buishand-2008a}, and \citet{blanchet-2011}.

Because closed-form expressions for the likelihoods associated with
spatial max-stable processes are not available, parameter estimation
and inference is problematic. Taking advantage of the availability of
bivariate densities, \citet{padoan-2010a} suggest maximum pairwise
likelihood estimation and asymptotic inference based on a sandwich
matrix (composed of expected derivatives of the composite likelihood
function) to properly account for using a pairwise likelihood when
computing standard errors [\citet{godambe-1987a}]. Recently, \citet
{genton-2011} extend this approach using composite likelihood based on
trivariate densities. The problem of spatial prediction, conditional on
observations, for max-stable random fields (analogous to Kriging for
Gaussian processes) has also proven difficult. The recent conditional
sampling algorithm of \citet{wang-2010a} is capable of producing both
predictions and prediction standard errors for most spatial max-stable
models of practical interest, subject to discretization errors that can
be made arbitrarily small.

Bayesian estimation and inference for max-stable process models for
spatial data on a continuous domain has been elusive. Implementing
these models in a fully-Bayesian framework has several advantages,
including incorporation of prior information and natural uncertainty
assessment for model parameters and predictions. Approximate Bayesian
methods based on asymptotic properties of the pairwise likelihood
function are possible. \citet{ribatet-2010a} use an estimated sandwich
matrix to adjust the Metropolis ratio within an MCMC sampler, while
\citet{shaby-2010c} rotates and scales the MCMC sample post-hoc and
\citet
{smith-2009a} use pairwise likelihoods without adjustment. Bayesian
models that are not based on max stable processes have been used for
analysis of extreme values with spatial structure. \citet{cooley-2007a}
uses a hierarchical model with a conditionally-independent generalized
Pareto likelihood, incorporating all spatial dependence through
Gaussian process priors on the generalized Pareto likelihood
parameters. Spatial dependence has also been achieved through Bayesian
Gaussian copula models [\citet{sang-2010a}] and through a more flexible
copula based on a Dirichlet process construction [\citet{fuentes-2010a}].

We develop a new hierarchical Bayesian model for analyzing max-stable
processes. The responses are modeled\vadjust{\goodbreak} as independent univariate GEV
conditioned on spatial random effects with positive stable random
effect distribution. Positive stable random effects have been used to
model multivariate extremes with finite dimensions [\citet
{fougeres-2009,stephenson-2009}]. We extend this approach to
accommodate data on a continuous spatial domain. We show that the
resulting model is max-stable marginally over the random effects, and
that a limiting case of this construction provides a finite-dimensional
approximation to the well-known Gaussian extreme value process (GEVP)
of \citet{smith-1990a}, often referred to as the ``Smith process.''
Lower-dimensional representations have previously been proposed for
high-dimensional extremes in various settings [\citet
{pickands-1981,Deheuvel-1983,Schlather-2002,Ehlert-2008,wang-2010a,wang-2010c,oesting-2011a,engelke-2011a}].
Our construction permits analysis of the joint distribution of all
observations, and thus can produce straightforward predictions at
unobserved locations.
Because we use a hierarchical model to represent the spatial max-stable
process, a Bayesian implementation is a natural choice. This allows us
to model underlying marginal structures as flexibly as we like, in
addition to automatic pooling of information and uncertainty
propagation. Also, the proposed framework permits representing the the
spatial process using a lower-dimensional representation, which leads
to efficient computing for large spatial data sets.

The remainder of the paper proceeds as follows. Section~\ref{smodel}
describes the model, which is compared to the GEVP in Section~\ref
{ssmith}. The method is evaluated using a simulation study in Section~\ref{ssim}.
In Section~\ref{stemp} we use the proposed method to
analyze yearly maximum precipitation using regional climate model
output from the North American Regional Climate Change Assessment
Program (NARCCAP) in the eastern US. Section~\ref{sconc} concludes.

\section{The hierarchical max-stable process model}\label{smodel}
\subsection{Spatial random effects model}
Let $Y(\bs)$ be the extreme value at location~$\bs$, defined over the
region $\bs\in\calD\subset\calR^2$. Here we describe a max-stable model
for $Y(\bs)$ assuming that it is a block-maximum, that is, the maximum
of many observations taken at location $\bs$, such as the yearly
maximum of daily precipitation levels. However, we note that max-stable
models are increasingly being used to model extreme individual
observations using a points above threshold approach [\citet
{Huser-2012}], and the residual max-stable process model described here
may be applicable to this type of analysis as well. We describe a model
for a single realization of the process and extend to multiple
independent realizations in Section~\ref{sST}.

Assuming the process is max-stable, then the marginal distribution of
$Y(\bs)$ is $\operatorname{GEV}[\mu(\bs),\sigma(\bs),\xi(\bs)]$, where $\mu
(\bs)$ is
the location, $\sigma(\bs)>0$ is the scale, and $\xi(\bs)$ is the shape
(GEV distribution is described in Appendix~\ref{appa1}). Equivalently [\citet
{resnick-1987}], we may express $Y(\bs) = \mu(\bs) + \frac{\sigma
(\bs
)}{\xi(\bs)} [X(\bs)^{\xi(\bs)}-1 ]$, where $X(\bs)$
is the
residual max-stable process with unit $\Frechet$ margins, that is,
$X(\bs)\sim \operatorname{GEV}(1,1,1)$. To allow for both nonspatial and spatial
residual variability, we model $X(\bs)$ as the product $X(\bs)=U(\bs
)\theta(\bs)$. Borrowing a term from geostatistics, we refer to
$U(\bs
)$ as the nugget effect since it accounts for nonspatial variation due
to measurement error or other small-scale features. The nugget is
modeled as $U(\bs)\iid \operatorname{GEV}(1,\alpha,\alpha)$, where, as described in
detail below, the parameter $\alpha\in(0,1)$ controls the relative
contribution to the nugget effect.

Residual spatial dependence is captured by $\theta(\bs)$. We express
the spatial process as a function of a linear combination of $L$ kernel
basis functions $w_l(\bs)\ge0$, scaled so that $\sum_{l=1}^Lw_l(\bs
)=1$ for all $\bs$. The spatial process is $\theta(\bs)= [\sum_{l=1}^LA_{l}w_{l}(\bs)^{1/\alpha} ]^{\alpha}$, where $A_l$ are
the basis function coefficients. To ensure max-stability and $\Frechet$
marginal distributions, the random effects $A_l$ follow the positive
stable distribution with density $p(A|\alpha)$ which has Laplace
transformation $\int_0^\infty\exp(-At)p(A|\alpha)\,dA = \exp
(-t^\alpha
)$. We denote this as $A_l\sim\operatorname{PS}(\alpha)$. Although $p(A|\alpha)$
has no closed form, it possesses the essential feature that if
$A_1,\ldots,A_T\iid \operatorname{PS}(\alpha)$, then $(A_1+\cdots+A_T)/T^{1/\alpha}\sim \operatorname{PS}(\alpha)$.
Appendix~\ref{appa2} verifies that this model for $X(\bs)$ is
max-stable with unit $\Frechet$ marginal distributions.

Marginalizing over the nugget terms $U(\bs)$ gives the hierarchical model
%
\begin{eqnarray}
\label{auxmodel} Y(\bs_i) | A_{1},\ldots,A_{L} &
\indep& \operatorname{GEV}\bigl[\mu^*(\bs_i),\sigma^*(
\bs_i),\xi^*(\bs_i)\bigr],
\nonumber
\\[-8pt]
\\[-8pt]
\nonumber
A_{l} &\iid& \operatorname{PS}(\alpha),
\end{eqnarray}
where $\mu^*(\bs) = \mu(\bs) + \frac{\sigma(\bs)}{\xi(\bs
)} [\theta
(\bs)^{\xi(\bs)}-1 ]$, $\sigma^*(\bs) = \alpha\sigma(\bs
)\theta(\bs
)^{\xi(\bs)}$, and
$\xi^*(\bs)=\alpha\xi(\bs)$. The responses are conditionally
independent given the random effects~$\bA$. The effect of conditioning
on $\bA= (A_{1},\ldots,A_{L})^T$, and thus the spatial process $\theta$,
is to move spatial dependence from the residuals to a random effect in
the GEV parameters. Marginalizing over the random effects induces
spatial dependence. The joint distribution function of the residual
process $X$ at $n$ locations $\bs_1, \ldots, \bs_n$ is
%
\begin{equation}
\label{ASL} \mathrm{P}\bigl(X(\bs_i)<c_i, i=1,\ldots,n
\bigr) = \exp \Biggl\{-\sum_{l=1}^L \Biggl[
\sum_{i=1}^n \biggl(\frac{w_l(\bs_i)}{c_i}
\biggr)^{1/\alpha} \Biggr]^{\alpha
} \Biggr\}.
\end{equation}
Therefore, although this is a process model defined on a continuous
spatial domain, the finite-dimensional distributions are multivariate
GEV (MGEV) with asymmetric logistic dependence function [\citet{Tawn-1990}].

Spatial dependence is often summarized by the extremal coefficient
[\citet{smith-1990a}]. The pairwise extremal coefficient $\vartheta
(\bs_i,\bs_j) \in[1,2]$ is defined by the relationship
%
\begin{equation}
\label{ECdef} P\bigl(X(\bs_i)<c,X(\bs_j)<c\bigr) = P
\bigl(X(\bs_i)<c\bigr)^{\vartheta(\bs_i,\bs_j)}.
\end{equation}
If $X(\bs_i)$ and $X(\bs_j)$ are independent, then $\vartheta(\bs_i,\bs_j) = 2$; in contrast, if $X(\bs_i)$ and $X(\bs_j)$ are completely
dependent, then $\vartheta(\bs_i,\bs_j) = 1$. The extremal coefficient
introduced by (\ref{auxmodel}) is
%
\begin{equation}
\label{ECgeneral} \vartheta(\bs_i,\bs_j) = \sum
_{l=1}^L \bigl(w_{l}(
\bs_i)^{1/\alpha
}+w_{l}(\bs_j)^{1/\alpha}
\bigr)^{\alpha}.
\end{equation}
Therefore, the extremal coefficient is the sum (over the $L$ kernels)
of the $L^{1/\alpha}$ norms of the vectors $[w_l(\bs_i),w_l(\bs_j)]$.

\begin{figure}

\includegraphics{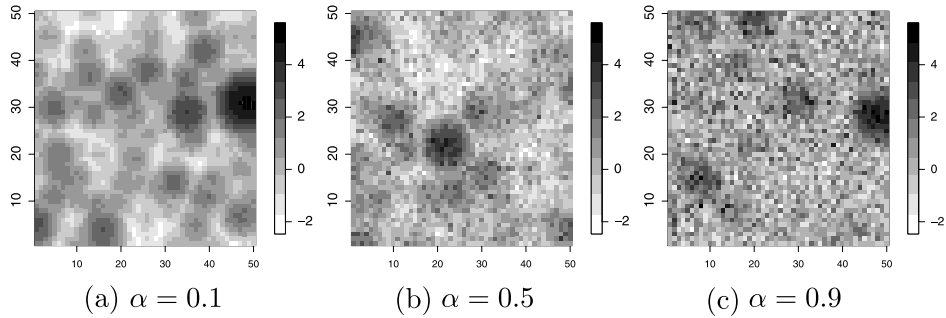}

\caption{Random draws on a $50\times50$ grid from the random effect
model for various $\alpha$.
Each sample has a knot at each data point, GEV parameters $\mu(\bs
)=\log
[\sigma(\bs)]=0$ and $\xi(\bs)=-0.1$,
and bandwidth $\tau=2$.}\label{fsamples}
\end{figure}

To see how $\alpha$ controls the nugget effect, consider two
observations at the same location, $\bs_i=\bs_j$. The two observations
share the same kernels, $w_l(\bs_i)=w_l(\bs_j)$ and thus $\theta(\bs_i)=\theta(\bs_j)$, but have different nugget terms $U(\bs_i)\ne
U(\bs_j)$. In this case, the extremal coefficient is $2^\alpha$. If $\alpha
=1$, then the nugget dominates and \mbox{$\vartheta(\bs_i,\bs_j)=2$} for all
pairs of locations, regardless of their spatial locations [since $\sum_{l=1}^Lw_l(\bs)=1$ for all $\bs$]. If $\alpha=0$, then $\vartheta
(\bs_i,\bs_j)=1$ when $\bs_i=\bs_j$, and there is no nugget effect. The
characteristics of the model are shown graphically in Figure~\ref
{fsamples}. In these random draws from the model, we see the process is
very smooth for $\alpha=0.1$ and has little discernable spatial pattern
with $\alpha= 0.9$.

The parameter $\alpha$ clearly plays an important role in this model.
It determines the magnitude of the nugget effect, the form of spatial
dependence function in (\ref{ECgeneral}), and the shape and scale of
the conditional distributions in (\ref{auxmodel}). To illustrate the
links between the contribution of $\alpha$ to these aspects of the
model, we consider the extreme cases with $\alpha=0$ and $\alpha=1$.
With $\alpha= 1$, $p(A|\alpha)$ concentrates its mass on $A=1$, and
thus $\theta(\bs)= \sum_{l=1}^Lw_{l}(\bs)=1$. In this case, the
conditional and marginal GEV parameters are the same, for example, $\mu^*(\bs)= \mu(\bs)$,
there is no residual dependence with $\theta
(\bs_i,\bs_j)=2$, and thus $Y(\bs)\indep\operatorname{GEV}[\mu(\bs),\sigma(\bs
),\xi(\bs
)]$. On the other hand, if $\alpha\approx0$, then the conditional
scale $\sigma^*(\bs)\approx0$ and $Y(\bs) \approx\mu^*(\bs)$,
a~continuous spatial process with strong small-scale spatial dependence
$\theta(\bs,\bs)\approx1$.

Spatial prediction (analogous to Kriging) at a new location $\bs^*$ is
straight-forward under this hierarchical model. Predictions are made by
simply computing $\theta(\bs^*)=[\sum_{l=1}^LA_{l}w_{l}(\bs^*)^{1/\alpha
}]^{\alpha}$, and then sampling $Y(\bs^*)$ from the independent GEV in
(\ref{auxmodel}). Repeating this at every MCMC iteration gives samples
from the posterior predictive distribution of $Y(\bs^*)$.

\subsection{Kernel and knot selection}\label{knot}

Although other kernels are possible, we use a scaled version of the
Gaussian kernel
%
\begin{equation}
\label{K} K(\bs|\bv_l,\tau) = \frac{1}{2\pi\tau^2}\exp \biggl[-
\frac
{1}{2\tau^2}(\bs-\bv_l)^T(\bs-\bv_l)
\biggr],
\end{equation}
where $\bv_1,\ldots,\bv_L\in\calR^2$ are spatial knots and $\tau>0$
is the
kernel bandwidth. To ensure that the kernels sum to one at each
location, the kernels are scaled as
%
\begin{equation}
\label{w} w_l(\bs) = \frac{K(\bs|\bv_{l},\tau)}{\sum_{j=1}^L K(\bs|\bv_{j},\tau)}.
\end{equation}
The knots are taken as a fixed and regularly-spaced grid of points
covering the spatial domain. Section~\ref{ssmith} shows that this
choice of kernel function and knot distribution gives the GEVP as a
limiting case. Even with a regular grid of knots, the extremal
coefficient is nonstationary, that is, $\vartheta(\bs_i,\bs_j)$ is not
simply a function of $\Vert \bs_i-\bs_j\Vert $. For example, $w(\bs_i)$ may not
equal $w(\bs_j)$ if $\bs_i$ is close to a knot and $\bs_j$ is not. This
discretization artifact dissipates for large $L$.

\begin{figure}

\includegraphics{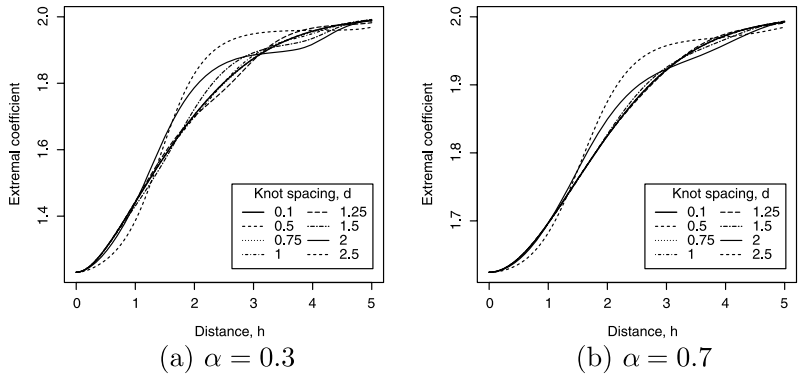}

\caption{Extremal coefficient $\vartheta(\bs_1,\bs_2)$, where $\bs_1 =
(0,0)$, $\bs_2=(0,h)$,
the kernel bandwidth is $\tau=1$, and the knots form a rectangular grid
with grid spacing $d$.}\label{fextremal}
\end{figure}

While the extremal coefficient does not fully characterize spatial
dependence, it is useful for guiding knot selection. Knot selection
poses a trade-off between computational burden with too many knots and
poor fit with too few knots. Consider the case of a Gaussian kernel
with bandwidth $\tau=1$ and knots on a large rectangular grid with grid
spacing $d$. Figure~\ref{fextremal} plots the extremal coefficient for
points $(0,0)$ and $(0,h)$ as a function of separation distance $h$.
The extremal coefficient has nearly an identical shape for all $d$ less
than or equal to $\tau$. For $d=1.25\tau$, the extremal coefficient
differs slightly from the fine grids, and for $d>1.25\tau$ the extremal
coefficient deviates considerably from the fine grids, especially for
small~$\alpha$. These results will scale for other $\tau$, therefore, a
rule of thumb is to select the knots so that the grid spacing is
approximately equal to the kernel bandwidth. Knot selection is
discussed further in Section~\ref{ssim}.

\subsection{Adaption for the NARCCAP data}\label{sST}

In Section~\ref{stemp} we analyze climate model output for $T>1$ years,
which requires additional notation. Denote $Y_t(\bs)$ as the response
for year $t$ and site $\bs$. Assuming the years are independent and
identically distributed (over years, not space) gives
%
\begin{eqnarray}
\label{auxmodel2} Y_t(\bs_i) | A_{1t},\ldots,A_{Lt}
&\indep& \operatorname{GEV} \bigl[\mu_t^*(\bs_i),
\sigma_t^*(\bs_i),\xi^*(\bs_i) \bigr],
\nonumber
\\[-8pt]
\\[-8pt]
\nonumber
A_{lt} &\iid& \operatorname{PS}(\alpha),
\end{eqnarray}
where $\theta_t(\bs)=[\sum_{l=1}^LA_{lt}w_{l}(\bs)^{1/\alpha
}]^{\alpha
}$ is the spatial random effect for year $t$, $\mu_t^*(\bs) = \mu
(\bs)
+ \frac{\sigma(\bs)}{\xi(\bs)} (\theta_t(\bs)^{\xi(\bs
)}-1 )$,
$\sigma_t^*(\bs)=\alpha\sigma(\bs)\theta_t(\bs)^{\xi(\bs)}$,
and $\xi^*(\bs)=\alpha\xi(\bs)$. Note that while the GEV parameters conditioned
on $\theta_t(\bs)$ in (\ref{auxmodel2}) vary by year, marginally,
$Y_t(\bs)\sim\operatorname{GEV}[\mu(\bs),\sigma(\bs),\xi(\bs)]$ for all $t$.

Gaussian process priors are used for the GEV parameters $\mu(\bs)$,
$\gamma(\bs) = \log[\sigma(\bs)]$, and $\xi(\bs)$. The Gaussian process
$\mu$ has mean $\mathbf{x}(\bs)^T\bbeta_{\mu}$, where $\mathbf{x}(\bs)$ includes the
spatial covariates such as elevation. The spatial covariance of $\mu$
is $\Matern$ [\citet{Banerjee03,Cressie93,Handbook}] with variance
$\delta_{\mu}^2>0$, range $\rho_{\mu}>0$, and smoothness $\nu_{\mu}>0$.
The other GEV parameters $\gamma(\bs)$ and $\xi(\bs)$ are modeled
similarly. In some applications, it may also be desirable to allow for
the GEV parameters to evolve over time, perhaps following a separate
linear time trend at each location, which would be a straightforward
modification of this model. The MCMC algorithm used to sample from this
model is described in Appendix~\ref{appa3}.

\section{Connection with the Gaussian extreme value process}\label{ssmith}

The GEVP of \citet{smith-1990a} is a well-known spatial max-stable
process. In this section we show that the proposed positive stable
random effects model in Section~\ref{smodel} contains this model as a
limiting case. The GEVP\vadjust{\goodbreak} construction for the residual process is
%
\begin{equation}
\label{Smith} X(\bs) = \max\bigl\{h_{1}K(\bs|\bu_{1},
\Sigma),h_{2}K(\bs|\bu_{2},\Sigma ),\ldots\bigr\},
\end{equation}
where $\{(h_1,\bu_1), (h_2,\bu_2),\ldots\}$ follows a Poisson process with
intensity $\lambda(h,\bu) = h^{-2}I(h>0)$, and $K$ is a kernel function
standardized so that $\int K(\bs|\bu,\tau)\,d\bu=1$ for all $\bs$. The
construction (\ref{Smith}) is a special case of the de Haan [\citet
{dehaan-1984a}] spectral representation. A useful analogy is to think
of $X(\bs)$ as the maximum rainfall at site $\bs$, generated as the
maximum over a countably-infinite number of storms. The $k${th} storm
has center $\bu_{k}\in\calR^2$, intensity $h_{k}>0$, and spatial range
given by $K(\bs|\bu_{k},\tau)$.

Under this model, the joint distribution at locations $\bs_1,\ldots,\bs_n$ is
%
\begin{equation}
\label{joint1}
\quad\mathrm{P} \bigl[X(\bs_1)<c_1,\ldots,X(
\bs_n)<c_n \bigr] = \exp \biggl[-\int\max_i
\biggl\{\frac{K(\bs_i|\bu,\Sigma
)}{c_i} \biggr\}\,d\bu \biggr].
\end{equation}
The GEVP has extremal coefficient
%
\begin{equation}
\label{ECsmith} \vartheta(\bs_i,\bs_j) = \int\max \bigl
\{K(\bs_i|\bu,\Sigma ),K(\bs_j|\bu ,\Sigma) \bigr\}\,d\bu,
\end{equation}
which simplifies to $\vartheta(\bs_i,\bs_j)=2\Phi (\frac
{\Vert \bs_i-\bs_j\Vert }{2\tau} )$ for the Gaussian kernel (\ref{K}). This does not
include a nugget effect, since $\vartheta(\bs_i,\bs_j)=1$ if $\Vert \bs_i-\bs_j\Vert =0$.

The connection to the model in Section~\ref{smodel} is made by
restricting the storm locations to the set of $L$ knot locations $\{\bv_1,\ldots,\bv_L\}$ and
rescaling the kernels to sum to one as in (\ref
{w}), giving
%
\begin{equation}
\label{Smith3} X(\bs) = \max \bigl\{h_{1}w_1(
\bs),\ldots,h_{L}w_L(\bs) \bigr\}.
\end{equation}
This amounts to truncating the de Haan spectral representation. If
$h_l\sim  \operatorname{GEV}(1,1,1)$, then $X(\bs)$ is max-stable with joint distribution
%
\begin{equation}
\label{joint2} \mathrm{P} \bigl[X(\bs_1)<c_1,\ldots,X(
\bs_n)<c_n \bigr] = \exp \Biggl[-\sum
_{l=1}^L\max_i \biggl\{
\frac{w_l(\bs_i)}{c_i} \biggr\} \Biggr],
\end{equation}
which implies that the marginal distributions are unit $\Frechet$. For
equally-spaced knots, this distribution converges weakly to the full
GEVP distribution function~(\ref{joint1}) as $L$ increases. We note
that this finite approximation could be applied to other max-stable
models such as those in \citet{schlather-2002a} and \citet
{kabluchko-2009a} by allowing the functions $K$ to be suitably scaled
Gaussian processes, unlike the current approach where $K$ is a kernel function.

Using the model described by (\ref{joint2}) directly is problematic
because it may not yield a proper likelihood. The process (\ref
{Smith3}) at $n$ locations $\{X(\bs_1),\ldots,\break X(\bs_n)\}$ is completely
determined by the intensities $\{h_1,\ldots,h_L\}$. Therefore, the
likelihood for $\{X(\bs_1),\ldots,X(\bs_n)\}$ requires a map from $\{
X(\bs_1),\ldots,X(\bs_n)\}$ to $\{h_1,\ldots,h_L\}$. This map may not exist, for
example, if $L<n$, and generally does not have a closed form. This is
common in dimension reduction methods for Gaussian process models [e.g.,
\citet{higdon-1998}, \citet{banerjee-2008a}, and \citet
{cressie-2008a}].

As with the Gaussian process dimension reduction methods, the model in
Section~\ref{smodel} includes both a spatial process ($\theta$) and a
nonspatial nugget term ($U$). Comparing (\ref{ASL}) and (\ref{joint2}),
we see a result of the nugget effect is that the $L^{\infty}$ norm (the
maximum) in (\ref{joint2}) is replaced with the $L^{1/\alpha}$ norm,
and that (\ref{ASL}) converges weakly to (\ref{joint2}) as $\alpha$
goes to zero. Including a nugget aids in computation, as the likelihood
becomes a simple product of univariate GEV densities. Including a
nugget term also has advantages beyond computation. The GEVP has been
criticized as unrealistically smooth [\citet{blanchet-2011}], and so a
nugget may improve fit. Analogously, in the geostatistical literature
for Gaussian data a nugget is not required, but is used routinely to
account for small-scale phenomena that cannot be captured with a smooth
spatial process [\citet{Cressie93,Banerjee03,Handbook}].

\section{Simulation study}\label{ssim}

In this section we conduct a simulation study to verify that the MCMC
algorithm produces reliable results, to investigate sensitivity to knot
selection, and to determine which parameters are the most difficult to
estimate. Data and knots are placed on $m\times m$ regular grids
covering $[l,u]\times[l,u]$, denoted $\calS(m,l,u)$. For each
simulation design, we generate data from the model described in Section~\ref{sST}
at the $n=49$ locations $\calS(7,0,6)$ and $T=10$ independent
years. The GEV location parameter varies by site following the Gaussian
process with mean zero, variance one, and exponential spatial
correlation $\exp(-\Vert \bs_i-\bs_j\Vert /2)$. Unlike the analysis of the
NARCCAP data in Section~\ref{stemp}, the GEV scale and shape parameters
are assumed to be the same for all sites and fixed at $\sigma(\bs)=1$
and $\xi(\bs)=0.2$. We fix these parameters in the simulation study for
computational purposes, and because these spatially-varying parameters
will likely be hard to estimate for these moderately-sized simulated
data sets. The simulations vary by the nugget effect ($\alpha$), the
kernel bandwidth ($\tau$), and the number of knots used to generate the
data ($L_0$). The simulation designs are numbered:
\begin{longlist}[(1)]
\item[(1)]$L_0=49$ knots at $\calS(7,0,6)$, $\alpha=0.3$, $\tau=3$,
\item[(2)]$L_0=49$ knots at $\calS(7,0,6)$, $\alpha=0.7$, $\tau=3$,
\item[(3)]$L_0=25$ knots at $\calS(5,0,6)$, $\alpha=0.3$, $\tau=3$,
\item[(4)]$L_0=25$ knots at $\calS(5,0,6)$, $\alpha=0.7$, $\tau=3$,
\item[(5)]$L_0=10\mbox{,}000$ knots at $\calS(100,-1,7)$, $\alpha=0.4$, $\tau=1$.
\end{longlist}

For the first four designs, the number of knots used to
generate the data is small enough to permit fitting the model with the
correct number of knots. We use these examples to explore sensitivity
to knot selection. The final design with $L_0=10\mbox{,}000$ knots represents
the limiting case with more knots than can be fit computationally. Here
we fit several course grids of knots and compare performance as the
number of knots increases to provide recommendations on the number of
knots needed to provide a good approximation to the limiting process.

$M=50$ data sets are generated for each simulation design. For each
simulated data set, we fit the model with a varying number of knots.
For the first four designs we compare $L=25$ knots at $\calS(5,0,6)$
and $L=49$ knots at $\calS(7,0,6)$ to compare fits with the true knots
and either too few ($L=25$ for designs 1 and 2) or too many ($L=49$ for
designs 3 and 4) knots. For the final design we compare fits with 8
knot grids: $L=25$ knots at $\calS(5,-1,7), \ldots, L=144$ knots at
$\calS(12,-1,7)$. The spatial covariance parameters for the GEV
location have priors $\delta_{\mu}^2\sim \operatorname{InvGamma}(0.1,0.1)$ and range
$\rho_{\mu}\sim \operatorname{InvGamma}(0.1,0.1)$; for this relatively small spatial
domain we fix the smoothness parameter $\nu_{\mu}=0.5$, giving an
exponential covariance. The design matrix $\bX$ includes only the
intercept with $\beta_{\mu}\sim\mathrm{N}(0,100^2)$. The GEV log scale and
shape are constant across space and have $\mathrm{N}(0,1)$ and $\mathrm{N}(0,0.25^2)$
priors, respectively. The residual dependence parameters have priors
$\tau\sim \operatorname{InvGamma}(0.1,0.1)$ and $\alpha\sim \operatorname{Unif}(0,1)$.

\begin{figure}

\includegraphics{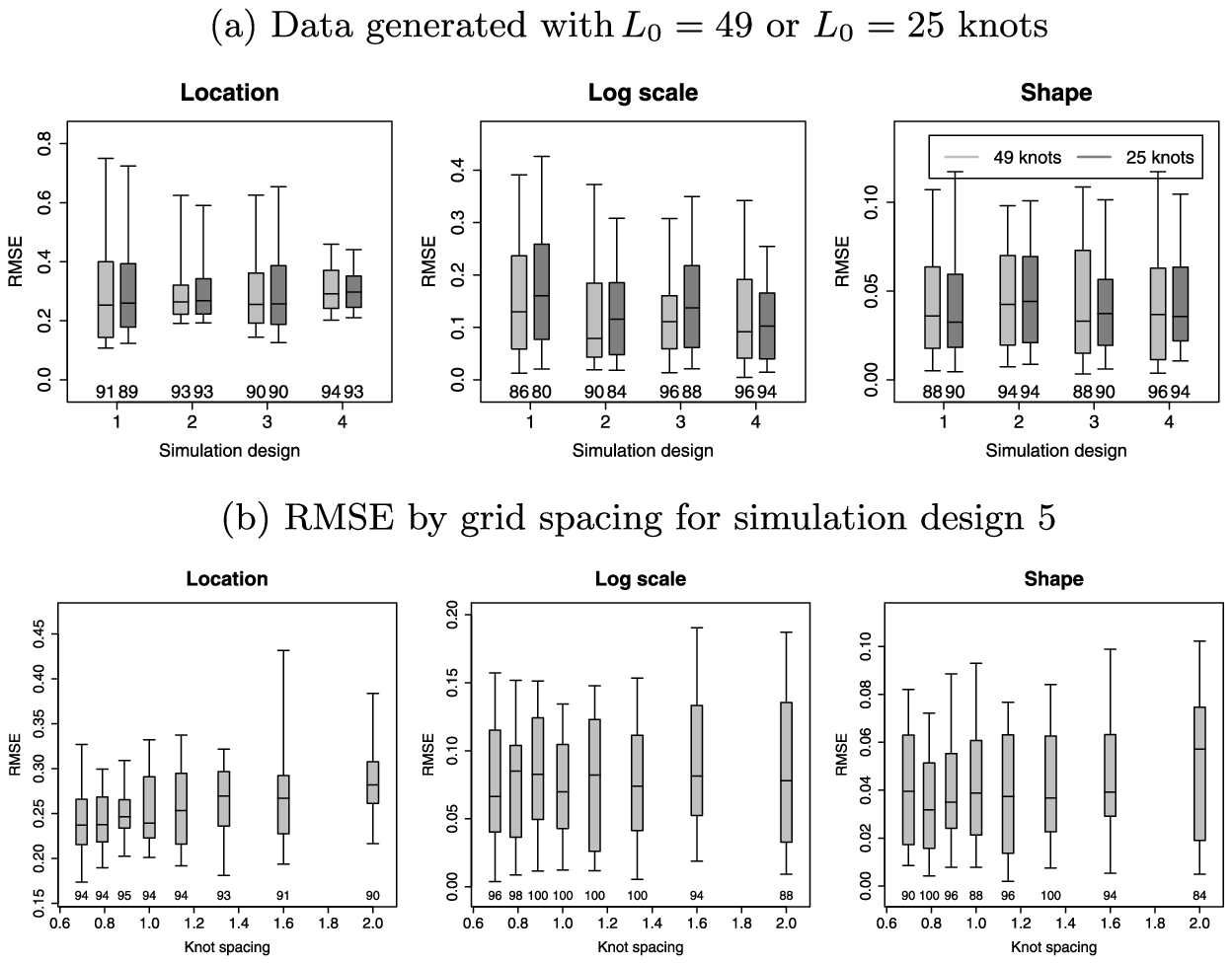}

\caption{Boxplots of root mean squared error (RMSE) for the GEV
parameters in the simulation study.
The horizontal lines in the boxplots are the 0.05, 0.25, 0.50, 0.75,
and 0.95 quantiles of RMSE.
Coverage percentages of posterior 95\% intervals are given below the
boxplots.}\label{fsim}
\end{figure}

The results are presented in Figure~\ref{fsim}. For each data set, we
compute the posterior mean of the GEV parameters at each location and
the mean squared error (MSE) of the posterior means (averaged over the
$n$ sites for the spatially-varying GEV location). Figure~\ref{fsim}
plots the $M=50$ root MSEs and coverage probabilities (averaged over
sites for the GEV location).

%
%
%

For the data generated with $L_0=25$ or $L_0=49$ knots in Figure~\ref
{fsim}(a), the coverage probabilities are generally near the nominal
level. With $L=49$ knots, the coverage probabilities range from 0.90 to
0.94 for the GEV location. For the first two designs, the model with
$L=25$ knots has fewer knots than were used to generate the data. This
does not have a substantial impact on the estimation of the GEV
location. However, using too few knots leads to increased RMSE and
under-coverage for the GEV log scale, especially for design 1 with
strong spatial dependence. For simulation designs 3 and 4, the model
with $L=49$ knots has nearly twice as many knots than were used to
generate the data. In these cases, the $L=49$ model performs nearly as
well as the correct $L=25$ model. For these simulation settings, we
conclude that using too few knots can lead to poor results, especially
for the scale parameter, but that including too many knots does not
degrade performance.

For the data generated with $L_0=10\mbox{,}000$ knots in Figure~\ref{fsim}(b),
we use knots grids with $L = 25,36,\ldots,144$ points. For comparison with
the kernel bandwidth, rather than plotting the results by $L$, we plot
results by the spacing between adjacent knots in the same column or
row, which ranges from 0.70 for $L=144$ to 2.00 for $L=25$. The
coverage probabilities are near or above the nominal level for all grid
spacings at or below the bandwidth, $\tau=1.0$, and the RMSE appears to
be fairly constant for all grid spacings at least as small as the
bandwidth. Therefore, this appears to be a reasonable rule of thumb for
selecting the number of knots.

We also computed RMSE for the spatial dependence parameters $\alpha$
and $\tau$ (not shown in Figure~\ref{fsim}) for this final case. For
$\alpha$, the average (over data sets) RMSE was 0.049 (coverage
percentage 96\%), 0.060 (88\%), and 0.101 (40\%) for grid spacings 0.7,
1.0, and 2.0, respectively. For $\tau$, the average RMSE was 0.102
(88\%
), 0.107 (90\%), and 0.233 (38\%) for grid spacings 0.7, 1.0, and 2.0,
respectively. As with the GEV parameters, the approximation with the
grid spacing at least as small as the bandwidth appears to provide
reasonable estimation of the spatial dependence parameters. When too
few knots are used, the bandwidth is often overestimated to compensate
for the lack of knots and, thus, RMSE is high and coverage is far below
the nominal level.\looseness=1\eject

\section{Analysis of regional climate model output}\label{stemp}

To illustrate the proposed method, we analyze climate model output
provided by the North American Regional Climate Change Assessment
Program (NARCCAP). Our objective is to study changes in extreme
precipitation under various climate scenarios in different spatial
regions while accounting for residual spatial dependence remaining
after allowing for spatially-varying GEV parameters. The data are
downloaded from the website \url{http://www.narccap.ucar.edu/index.html}. We analyze output from two
timeslice experiments. Both runs use the Geophysical Fluid Dynamics
Laboratory's AM2.1 climate model with 50~km resolution. The model is run
separately under historical (1969--2000) and future conditions
(2039--2070). Observational data is used for the sea-surface
temperature and ice boundary conditions in the historical run. The
boundary conditions for the future run are perturbations of the
historical boundary conditions. The amount of perturbation is based on
a lower resolution climate model. The perturbations assume the A2
emissions scenario [Nakicenovic et al. (\citeyear{Na00})], which increases CO$_2$
concentration levels from the current values of about 380~ppm to about
870~ppm by the end of the 21st century.

\begin{figure}[b]

\includegraphics{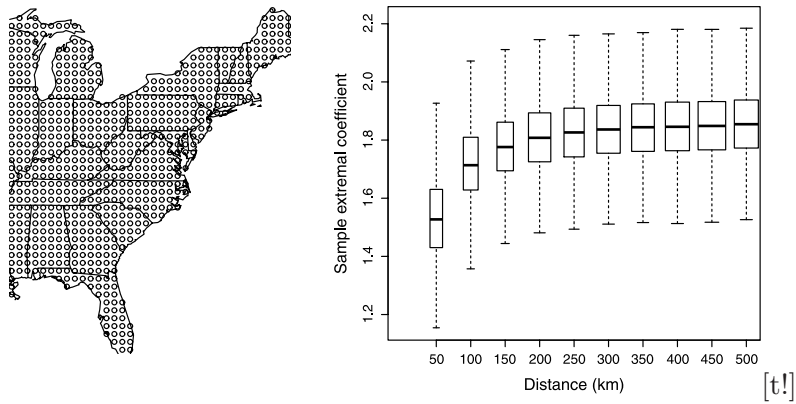}[t!]

\caption{Grid cell centers for the NARCCAP output (left) and madogram
extremal coefficient estimate
(box width is proportional to the number of observations) for the
historical run.}\label{fdata}
\end{figure}

We analyze data for $n=697$ grid cells in eastern US as shown in Figure~\ref{fdata}.
For grid cell $i$ with location $\bs_i$ and year $t$, we
take the annual maximum of the daily precipitation totals as the
response, $Y_{t}(\bs_i)$. NARCCAP provides eight 3-hour precipitation
rates each day, and we compute the daily total by summing these eight
values and multiplying by three. To explore the form of residual
spatial dependence, we use the madogram [\citet{cooley-2006a}] function
in the \texttt{SpatialExtremes} package in~$\mathtt{R}$ (\href{http://www.R-project.org}{www.r-project.org}).
The madogram converts the observations at each
site to have unit $\Frechet$ margins using a rank transformation, and
then estimates the pairwise extremal coefficients. Figure~\ref{fdata}
plots the estimated extremal coefficients against $\Vert \bs_i-\bs_j\Vert $.
This plot clearly shows residual spatial dependence.

The data from the two runs are analyzed separately using the model
described in Section~\ref{smodel}. We assume that the process is
stationary in time during each period, that is, the GEV marginal
density at each location is constant over time in each simulation
period. We use $n=L$ terms with knots at the data points $\bs_1,\ldots,\bs_n$.
The residual dependence parameters have priors
$\tau\sim \operatorname{InvGamma}(0.1,0.1)$ and $\alpha\sim \operatorname{Unif}(0,1)$. For both scenarios, all
three GEV parameters vary spatially following Gaussian process priors.
The covariates for the mean of the GEV parameters, $\mathbf{x}(\bs
)$, include
the intercept, grid cell latitude, longitude, elevation, and log
elevation. The elements of $\bbeta_{\mu}$ have independent $\mathrm{N}(0,100^2)$
priors. The spatial covariance parameters have priors $\delta_{j}^2\sim
 \operatorname{InvGamma}(0.1,0.1)$, range $\rho_{j}\sim \operatorname{InvGamma}(0.1,0.1$), and
smoothness $\nu_{j}\sim \operatorname{InvGamma}(0.1,0.1)$ for $j\in\{\mu,\gamma
,\xi\}$.

\begin{figure}

\includegraphics{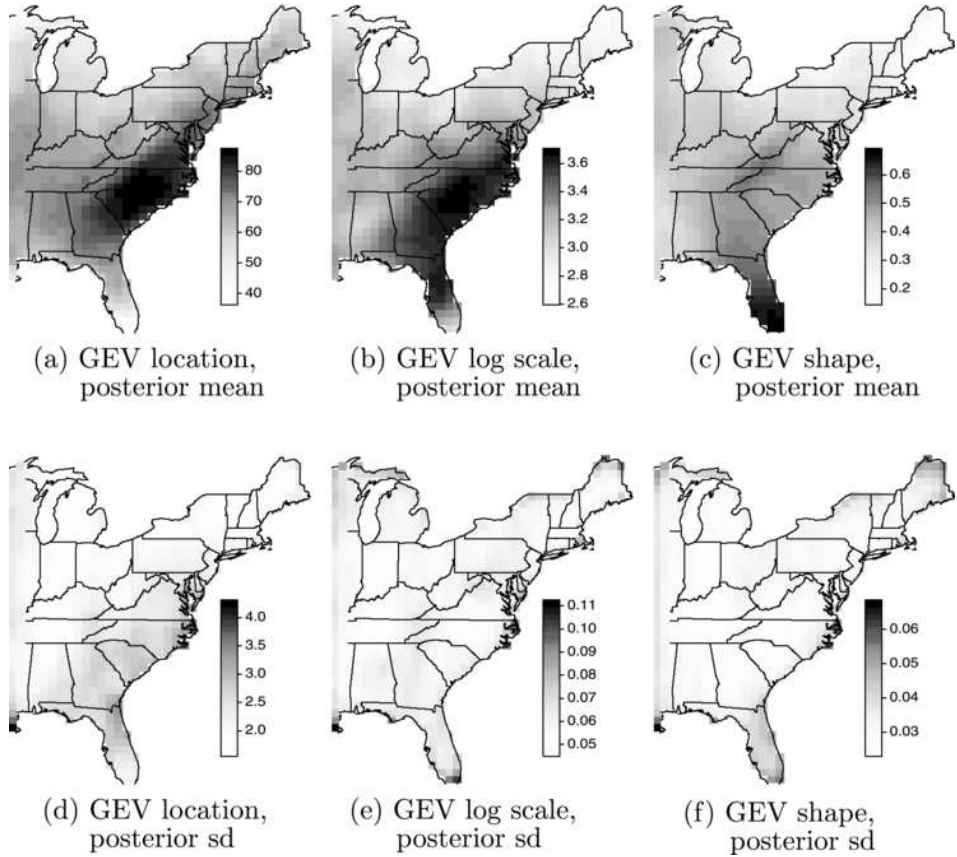}

\caption{Posterior mean and standard deviation of the GEV location, log
scale, and
shape parameters for the historical simulation. All units are
mm/h.}\label{fgevhist}
\end{figure}

\begin{figure}

\includegraphics{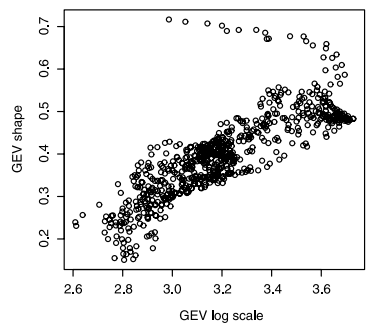}

\caption{Plot of the posterior mean GEV scale versus posterior mean GEV
shape at each site.}\label{fshapescale}
\end{figure}

Figure~\ref{fgevhist} shows the estimated GEV parameters for the
historical simulation. The estimated location and log scale parameters
are highest in the southeast. The posterior mean of the GEV shape is
generally positive, indicating a right-skewed distribution with no
upper bound. The estimated shape is the largest in Florida. Comparing
the posterior means and standard deviations, there is evidence that all
three GEV parameters vary spatially. Figure~\ref{fshapescale} shows
that there is strong positive dependence between the shape and scale as
one might expect, since for shape in $(0,0.5)$ both the mean and variance
of GEV includes the ratio of the scale and shape. For locations with
large shapes, there is a negative dependence with the log scale.

To formally assess the need for spatially-varying GEV parameters, we
also refit the model for the historical simulation using the Bayesian
variable selection prior of \citet{reich-2010} to test whether the
variance $\delta_{j}^2$ equals small constant $\Delta_j^2=0.01^2$. The
test is carried out using the mixture prior $\delta_{j} = g_j\delta_j^*
+ (1-g_j)\Delta_0$, where $g_j\sim \operatorname{Bernoulli}(0.5)$ and $\delta_j^{*2}\sim \operatorname{InvGamma}(0.1,0.1$).
The intuition behind this prior is that
if $g_j=1$, then $\delta_{j}^2\sim \operatorname{InvGamma}(0.1,0.1)$ and the GEV
parameter varies spatially; in contrast, if $g_0=0$, then $\delta_{j}^2=\Delta_j^2$,
and spatial variation after accounting for spatial
covariates $\mathbf{x}$ is negligible. Therefore, the posterior mean
of $g_j$
can be interpreted as the posterior probability that the $j${th} GEV
parameter varies spatially, which can be used to approximate the Bayes
factor comparing these models. In the separate mixture prior fit, the
posterior probability that the GEV parameters vary spatially was at
least 0.99 for all three parameters.

\begin{figure}

\includegraphics{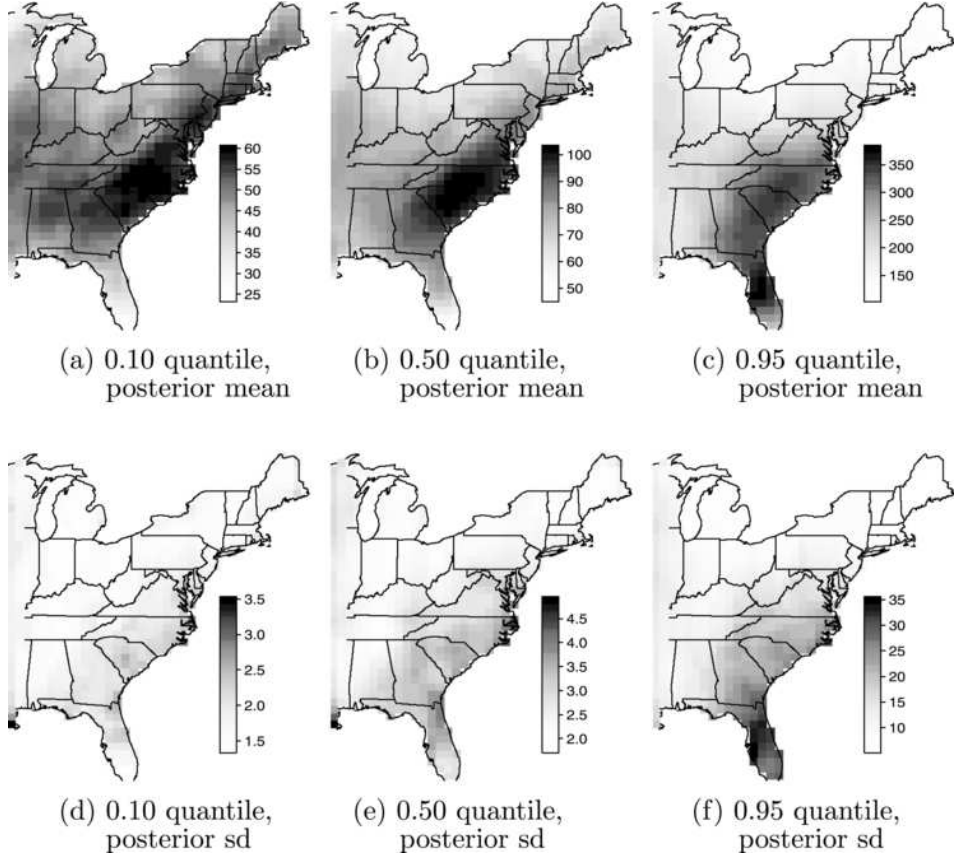}

\caption{Posterior mean and standard deviation of the 0.10, 0.50, and
0.95 quantiles
for the historical simulation. All units are mm/h.}\label{fquanthist}
\end{figure}

We also aim to quantify changes in extreme quantiles. The $q${th}
quantile at location $\bs$ is $\mu(\bs) + \sigma(\bs) [1-\log
(1/q)^{-\xi(\bs)} ]/\xi(\bs)$, which is also called the $1/(1-q)$
year return level. Figure~\ref{fquanthist} plots the posterior of
various pointwise quantile levels. The large location and scale
parameters lead to large medians in the southeast, while the 0.95
quantile is the largest in Florida due to the large shape parameter.

\begin{figure}

\includegraphics{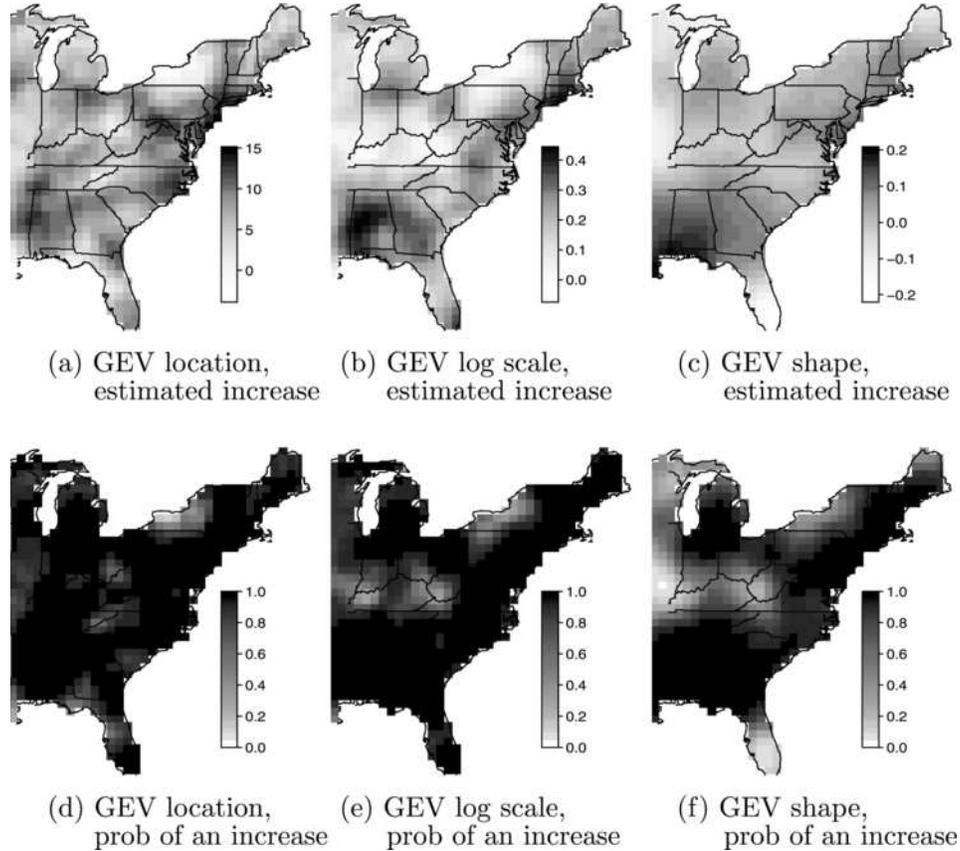}

\caption{Posterior mean change from historical to future time
simulation and the posterior
probability that this change is positive for the GEV location, log
scale, and shape parameters.
All units are mm/h.}\label{fgevchange}
\end{figure}

\begin{figure}

\includegraphics{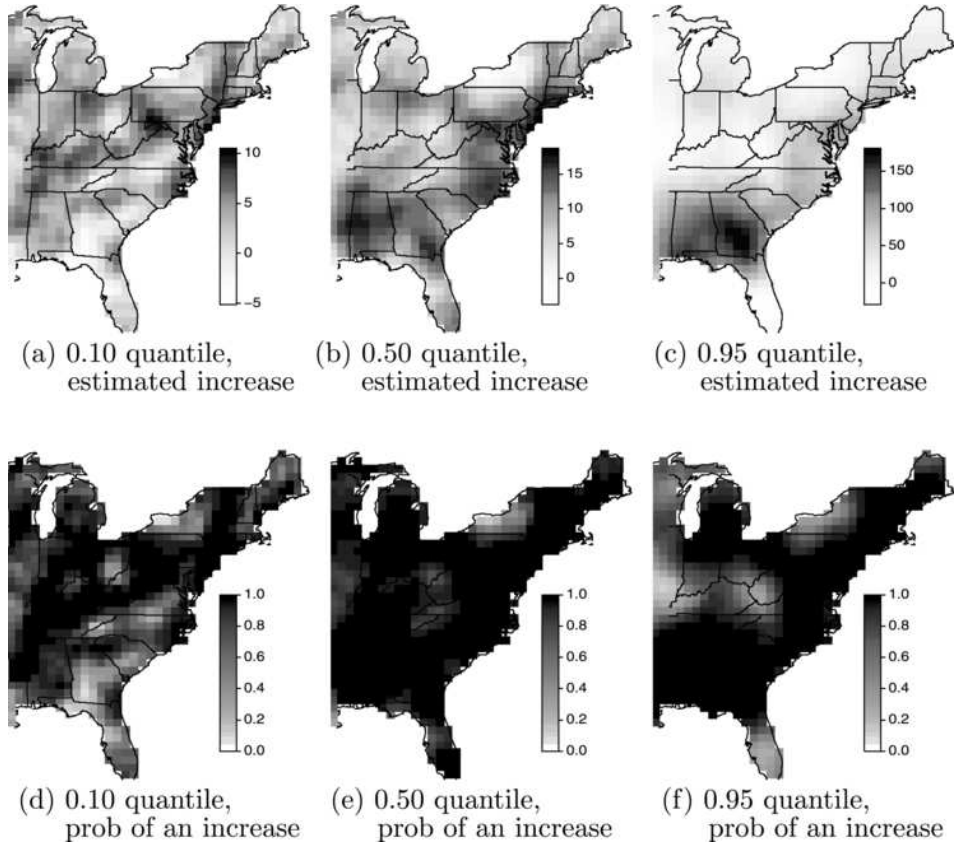}

\caption{Posterior mean change from historical to future time
simulations and the posterior probability
that this change is positive for the 0.10, 0.50, and 0.95 quantiles.
All units are mm/h.}\label{fquantchange}
\end{figure}

The difference between the historical and future scenarios is
summarized in Figures~\ref{fgevchange} and~\ref{fquantchange}. The
estimated GEV location and log scale parameters are larger for the
future scenario for the majority of the spatial domain. The increase is
the largest in Alabama, Georgia, and New England. The shape parameter
also shows an increase in Alabama, but statistically significant
decrease in Florida. Figure~\ref{fquantchange}(c) shows that these
changes in GEV parameters lead to an increase in the 0.95 quantile for
most of the spatial domain. With the exception of the midwest and
southern Florida, the posterior probability of an increase in the 0.95
quantile is near one [Figure~\ref{fquantchange}(d)], indicating that
extremes have a different spatial pattern in the future scenario.

\begin{figure}

\includegraphics{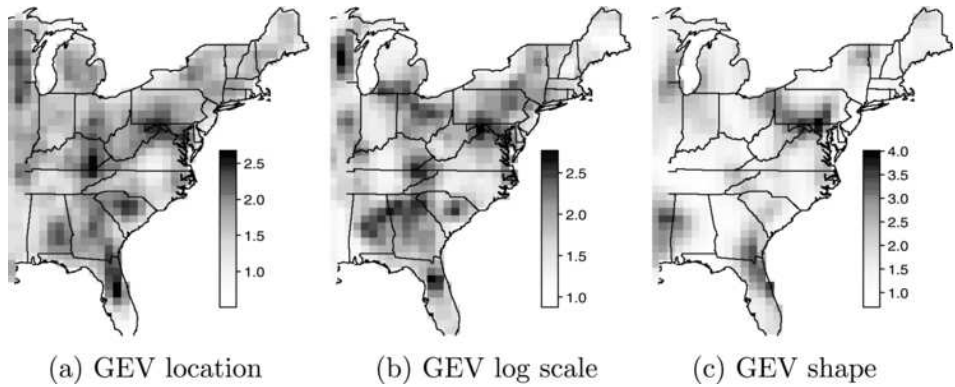}

\caption{Ratio of the posterior variance of the GEV parameters for the
models with and without residual spatial dependence.}\label{fvarratio}
\end{figure}

Parameter estimates provide evidence of residual dependence: the
posterior mean (standard deviation) of $\alpha$ is 0.483 (0.008) and
the posterior mean of the spatial range $\tau$ is 41.6 (0.4)
kilometers. To illustrate the effects of failing to account for
residual spatial dependence, we compare these results with the model
that ignores spatial dependence in the residuals, that is, sets $\alpha
=1$. One effect of accounting for residual dependence is an increase in
posterior variance for the GEV parameters. Figure~\ref{fvarratio} shows
that the posterior variance often doubles as a result of including
residual dependence. Therefore, while spatial modeling of the GEV
parameters reduces uncertainty by borrowing strength across space
compared to analyzing all sites completely separately, it appears that
spatial modeling of the GEV parameters without accounting for residual
dependence underestimates uncertainty.\looseness=1\eject

\section{Discussion}\label{sconc}

In this paper we propose a new modeling approach for spatial max-stable
processes. The proposed model is closely related to the GEVP and
permits a Bayesian analysis via MCMC methods. Applied to the climate
data, we find statistically significant increases under the future
climate scenario in the upper quantiles of precipitation for most of
the spatial domain, with the largest increase in the southeast.

The proposed hierarchical model opens the door for several exciting
research directions. The model could be made even more flexible by
changing the form of the kernels. It should be possible to replace the
Gaussian kernel with any other kernel that integrates to one, that is,
any other two-dimensional density function. For large data sets, it may
even be possible to estimate the kernel function nonparametrically from
the data. \citet{Zheng-2010} and \citet{reich-2011} use Bayesian
nonparametrics to estimate the spatial covariance function of a
Gaussian process. This approach could be extended to the extreme data,
using, say, a Dirichlet process mixture prior for the kernel function.
The methods proposed in this paper could also be extended to more
complicated dependency structures. For example, we have ignored the
temporal dependence because the spatial association is far stronger
than the temporal association for these data. However, using
three-dimensional kernels (two for space, one for time) would give a
feasible max-stable model for spatiotemporal data.

\begin{appendix}
\section*{Appendix}

\subsection{Generalized extreme value (GEV) distribution}\label{appa1}

The GEV distribution has three parameters: location $\mu$, scale
$\sigma
>0$, and shape $\xi$. If $Y\sim\operatorname{GEV}(\mu,\sigma,\xi)$, then $Y$ has
distribution function $P(Y<y) = \exp[-t(y)]$ and density $f(y) = \frac
{1}{\sigma}t(y)^{\xi+1}\exp[-t(y)]$, where
\[
t(y) = \cases{
\displaystyle\biggl[1+\frac{\xi}{\sigma}(y-
\mu) \biggr]^{-1/\xi}, &\quad $\xi\ne0,$
\vspace*{2pt}\cr
\displaystyle\exp\bigl[-(y-\mu)/\sigma\bigr], & \quad $\xi=0.$ }
\]
The shape parameter determines the support, with $Y\in(-\infty,\mu
-\sigma/\xi]$ if $\xi<0$, $Y\in(-\infty,\infty)$ is $\xi=0$, and
$Y\in
[\mu-\sigma/\xi,\infty)$ in $\xi>0$. The GEV has three well-known
subfamilies defined by the shape: the Weibull ($\xi<0$), Gumbel ($\xi
=0$), and $\Frechet$ ($\xi>0$) families.

\subsection{Properties of the random effects model}\label{appa2}

Here we show that the hierarchical representation in (\ref{auxmodel})
is max-stable and has GEV margins.

\textit{GEV marginal distributions}: Since the margins are identical for
all locations, we omit the notational dependence on $\bs$. The marginal
distribution function of $X$ is
%
\begin{eqnarray}
P(X<c) &=& \int P(X|\bA)p(\bA|\alpha)\,d\bA\nonumber\\
&=& \int\exp \biggl\{- \biggl[1+
\frac{\alpha}{\alpha\theta}(c-\theta) \biggr]^{-1/\alpha
} \biggr\}p(\bA |\alpha)\,d\bA
\nonumber\\
&=&\int\exp \Biggl\{-c^{-1/\alpha} \Biggl(\sum_{l=1}^LA_lw_l^{1/\alpha}
\Biggr) \Biggr\}p(\bA|\alpha)\,d\bA
\\
&=&\prod_{l=1}^L\int\exp \bigl
\{-c^{-1/\alpha}w_l^{1/\alpha
}A_l \bigr\}
p(A_l|\alpha)\,dA_l
\nonumber
\\
&=&\prod_{l=1}^L \exp \bigl\{-
\bigl(c^{-1/\alpha}w_l^{1/\alpha
} \bigr)^\alpha \bigr
\} =\exp \Biggl(-\frac{1}{c}\sum_{l=1}^Lw_l
\Biggr) =\exp (-1/c).
\nonumber
\end{eqnarray}
This is the unit $\Frechet$ distribution function.

\textit{Max-stability}: The process is max-stable if for any set of
locations $\{\bs_1,\ldots,\break\bs_n\}$ and any $t\,{>}\,0$, $\operatorname{Prob} [X(\bs_1)
\,{\leq}\,tc_1, \ldots, X(\bs_n)\,{\leq}\,tc_n ]^t\,{=}\,\operatorname{Prob} [X(\bs_1)
\,{\leq}\,c_1, \ldots,\break  X(\bs_n) \leq c_n ]$ [e.g., \citet{zhang-2010}].
From (\ref{ASL}),
\begin{eqnarray*}
&&\operatorname{Prob} \bigl[X(\bs_1) \leq tc_1, \ldots,
X(\bs_n) \leq tc_n \bigr]^t\\
&&\qquad= \exp \Biggl\{-
\sum_{l=1}^L \Biggl[\sum
_{i=1}^n \biggl(\frac{w_l(\bs_i)}{tc_i}
\biggr)^{1/\alpha} \Biggr]^{\alpha} \Biggr\}^t
\\
&&\qquad= \exp \Biggl\{-\frac{1}{t}\sum_{l=1}^L
\Biggl[\sum_{i=1}^n \biggl(
\frac
{w_l(\bs_i)}{c_i} \biggr)^{1/\alpha} \Biggr]^{\alpha} \Biggr
\}^t
\\
&&\qquad=\operatorname{Prob} \bigl[X(\bs_1) \leq c_1, \ldots,
X(\bs_n) \leq c_n \bigr].
\end{eqnarray*}

\subsection{MCMC details}\label{appa3}

A complication that arises when using positive stable random effects is
that their density does not have a closed form. To overcome this
problem, we use the auxiliary variable technique of \citet
{stephenson-2009} for the asymmetric logistic MGEV. \citet
{stephenson-2009} introduces auxiliary variables $B_{l}\in(0,1)$ so that
%
\begin{equation}
\label{psu} p(A,B|\alpha) = \frac{\alpha A^{-1/(1-\alpha)}}{1-\alpha}c(B)\exp \bigl[-c(B)A^{-\alpha/(1-\alpha)}
\bigr],
\end{equation}
where $c(B) =  [\frac{\sin(\alpha\pi B)}{\sin(\pi B)}
]^{1/(1-\alpha)}\frac{\sin[(1-\alpha) \pi B]}{\sin(\alpha\pi B)}$.
Then, marginally over $B_{l}$, $A_{l}\sim \operatorname{PS}(\alpha)$. This
marginalization is handled naturally via MCMC. Incorporating the
auxiliary variable gives
%
\begin{eqnarray}
\label{auxmodel3} Y_t(\bs_i) | A_{1t},B_{1t}\ldots,A_{Lt},B_{Lt}
&\indep& \operatorname{GEV} \bigl[\mu_t^*(\bs_i),
\sigma_t^*(\bs_i),\xi^*(\bs_i) \bigr],
\nonumber
\\[-8pt]
\\[-8pt]
\nonumber
(A_{lt},B_{lt}) &\iid& p(A,B|\alpha),
\nonumber
\end{eqnarray}
which is the model fit to the data.

We perform MCMC sampling for the model in (\ref{auxmodel3}) using R
(\texttt{\href{http://www.r-project.org/}{http://}
\href{http://www.r-project.org/}{www.r-project.org/}}). The Metropolis within Gibbs
algorithm is used to draw posterior samples. This begins with an
initial value for each model parameter, and then parameters are updated
one-at-a-time, conditionally on all other parameters. The GEV
parameters $\mu$, $\sigma=\exp(\gamma)$, and $\xi$, spatial dependence
parameters $\tau$ and $\alpha$, and auxiliary variables $(A_l,B_l)$ are
updated using Metropolis updates. To update the GEV location at site
$\bs_i$ for the $r${th} MCMC iteration, we generate a candidate using
a random walk Gaussian candidate distribution $\mu^{(c)}(\bs_i)\sim\mathrm{N}(\mu^{(r-1)}(\bs_i),s^2)$,
where $\mu^{(r-1)}(\bs_i)$ is the
value at
MCMC iteration $r-1$ and $s$ is a tuning parameter. The acceptance
ratio is
\begin{eqnarray*}
R&=& \biggl\{\frac{\prod_{t=1}^Tl[Y_t(\bs_i)|\mu^{(c)}(\bs_i), \exp
[\gamma
(\bs_i)], \xi(\bs_i), \theta_t(\bs_i)]} {
\prod_{t=1}^Tl[Y_t(\bs_i)|\mu^{(r-1)}(\bs_i), \exp[\gamma(\bs_i)], \xi
(\bs_i), \theta_t(\bs_i)]} \biggr\}\\
&&{}\times  \biggl\{\frac{p[\mu^{(c)}(\bs_i)|\mu(\bs_j), j\ne i]} {
p[\mu^{(r-1)}(\bs_i)|\mu(\bs_j), j\ne i]} \biggr\},
\end{eqnarray*}
which is a function of the GEV likelihood of $Y_t(\bs)$ in (\ref
{auxmodel2}), denoted as $l[Y_t(\bs)|\mu(\bs),  \exp[\gamma(\bs)],
\xi
(\bs), \theta_t(\bs)]$, as well as the full conditional prior of
$\mu
(\bs_i)$ given $\mu(\bs_j)$ for all $j\ne i$, $p[\mu(\bs_i)|\mu
(\bs_j),
j\ne i]$, which is found using the usual formula for the conditional
distribution of a multivariate normal. The candidate is accepted with
probability $\min\{R,1\}$. If the candidate is accepted, then $\mu^{(r)}(\bs_i)= \mu^{(c)}(\bs_i)$,
otherwise the previous value is
retained, $\mu^{(r)}(\bs_i)= \mu^{(r-1)}(\bs_i)$. The other GEV
parameters $\gamma(\bs_i)$ and $\xi(\bs_i)$ are updated similarly. GEV
hyperparameters, such as $\bbeta_{\mu}$ and spatial covariance
parameters, are updated conditionally on the GEV parameters and, thus,
their updates are identical to the usual Bayesian geostatistical model.

The spatial dependence parameters $\tau$ and $\alpha$ and the auxiliary
variables $A_{lt}$ and $B_{lt}$ are also updated using Metropolis
sampling. These updates differ from $\mu(\bs_i)$ only in their
acceptance ratios. For computing purposes, we transform to $\delta=
\log(\tau)$. The acceptance ratio for $\delta$ is
\[
\biggl\{\frac{\prod_{t=1}^T\prod_{i=1}^nl[Y_t(\bs_i)|\mu(\bs_i),
\exp
[\gamma(\bs_i)], \xi(\bs_i), \theta_t^{(c)}(\bs_i)]} {
\prod_{t=1}^T\prod_{i=1}^nl[Y_t(\bs_i)|\mu(\bs_i), \exp[\gamma
(\bs_i)], \xi(\bs_i), \theta_t^{(r-1)}(\bs_i)]} \biggr\} \biggl\{\frac{p[\delta^{(c)}]} {
p[\delta^{(r-1)}]} \biggr\},
\]
where $\theta^{(c)}_{t}$ and $\theta^{(r-1)}_{t}$ are the values of
$\theta_t$ evaluated with $\tau^{(c)}=\exp(\delta^{(c)})$ and $\tau^{(r-1)}=\exp(\delta^{(r-1)})$,
respectively, and $p(\delta)$ is the
log-gamma prior. The acceptance ratio for $\alpha$ is
\begin{eqnarray*}
&&\biggl\{\frac{\prod_{t=1}^T\prod_{i=1}^nl[Y_t(\bs_i)|\mu(\bs_i),
\exp
[\gamma(\bs_i)], \xi(\bs_i), \theta_t^{(c)}(\bs_i)]} {
\prod_{t=1}^T\prod_{i=1}^nl[Y_t(\bs_i)|\mu(\bs_i), \exp[\gamma
(\bs_i)], \xi(\bs_i), \theta_t^{(r-1)}(\bs_i)]} \biggr\} \\
&&\qquad{}\times \biggl\{\frac{\prod_{t=1}^T\prod_{l=1}^Lp(A_{lt},B_{lt}|\alpha^{(c)})} {
\prod_{t=1}^T\prod_{l=1}^Lp(A_{lt},B_{lt}|\alpha^{(r-1)})} \biggr\} I
\bigl(0<\alpha^{(c)}<1\bigr).
\end{eqnarray*}
We use a log-normal candidate distribution for $A_{lt}\sim$ LN$[\log
(A_{lt}^{(r-1)}),s_A^2]$, with density denoted $q(A^{(c)}_{lt}|A^{(r-1)}_{lt})$.
The latent variables $A_{tl}$ and $B_{lt}$ have acceptance ratios
\begin{eqnarray*}
&&\biggl\{\frac{\prod_{i=1}^nl[Y_t(\bs_i)|\mu(\bs_i), \exp[\gamma
(\bs_i)],
\xi(\bs_i), \theta_t^{(c)}(\bs_i)]} {
\prod_{i=1}^nl[Y_t(\bs_i)|\mu(\bs_i), \exp[\gamma(\bs_i)], \xi
(\bs_i), \theta_t^{(r-1)}(\bs_i)]} \biggr\}\\
&&\qquad{}\times  \biggl\{\frac{p(A^{(c)}_{lt},B_{lt}|\alpha)} {
p(A^{(r-1)}_{lt},B_{lt}|\alpha)} \biggr\} \biggl\{
\frac
{q(A^{(r-1)}_{lt}|A^{(c)}_{lt})}{q(A^{(c)}_{lt}|A^{(r-1)}_{lt})} \biggr\}
\end{eqnarray*}
for $A_{lt}$ and
\[
\frac{p(A_{lt},B^{(c)}_{lt}|\alpha)}{p(A_{lt},B^{(r-1)}_{lt}|\alpha
)}I\bigl(0<B_{lt}^{(c)}<1\bigr)
\]
for $B_{lt}$.

The standard deviations of all candidate distributions are adaptively
tuned during the burn-in period to give acceptance rates near 0.4. Note
that after the burn-in, the candidate distribution is fixed and this
defines a stationary Markov chain and satisfies the usual mixing
conditions, generating samples from the true posterior distribution
once convergence is reached. We generate two (one for the simulation
study) chains of length 25,000 samples and discard the first 10,000
samples of each chain as burn-in. Convergence is monitored using trace
plots and autocorrelation plots for several representative parameters.
\end{appendix}

\section*{Acknowledgments}
We also wish to acknowledge
several helpful discussions with Richard Smith of the University of
North Carolina---Chapel Hill and Alan Gelfand of Duke University.

%


\printaddresses

\end{document}